\RequirePackage{lineno}
\documentclass[aps,prl,showpacs,twocolumn,superscriptaddress,groupedaddress,floatfix,amsfonts]{revtex4}
\usepackage{amssymb}
\usepackage{amsmath}
\usepackage{xspace}
\usepackage{array}
\usepackage{epsfig}
\usepackage{psfrag}
\usepackage{graphicx}
\usepackage{dcolumn}
\usepackage{bm}
\usepackage{epsfig}
\usepackage{lineno}
\usepackage{multirow}
\usepackage{slashed}
\usepackage{rotating}
\usepackage{subfigure}
\usepackage{caption}
\usepackage{verbatim} 
\usepackage{mathrsfs} 

\newcommand{\pt}{\mbox{$p_{T}$}~}
 
\newcommand{\mjl}{\mbox{$D_{\text{MJL}}$}~}
\newcommand{\mjle}{\mbox{$D_{\text{MJL}}$}}

\newcommand{\dzero} {\mbox{D0}~}
\newcommand{\DO}{\mbox{D0}~}

\newcommand{\GeV} {\ensuremath{\mathrm{Ge\kern -0.1em V}}~}
\newcommand{\GeVe} {\ensuremath{\mathrm{Ge\kern -0.1em V}}}
\newcommand{\TeV} {\ensuremath{\mathrm{Te\kern -0.1em V}}}

\newcommand{\ppbar}{\mbox{$p\overline{p}$}~}

\newcommand{\ptj}{\mbox{$p_{T}^{\text{jet}}$}}

\newcommand{\etaj}{\mbox{$\eta^{\text{jet}}$}}

\newcommand{\ratio}{\mbox{$\sigma(Z+2~b~\text{jets})/\sigma(Z+\text{2~jets})$}}

\def\gsim{\mathrel{\rlap{\raise.4ex\hbox{$>$}} {\lower.6ex\hbox{$\sim$}}}}
\def\lsim{\mathrel{\rlap{\raise.4ex\hbox{$<$}} {\lower.6ex\hbox{$\sim$}}}}

\begin{document}

\hspace{5.2in} \mbox{FERMILAB-PUB-15-013-E}

\title{\boldmath Measurement of the ratio of inclusive cross sections\\ 
$\sigma (p\bar{p} \rightarrow Z+2~b~\text{jets}) / \sigma (p\bar{p} \rightarrow Z+ \text{2~jets})$ 
in $p\bar{p}$ collisions at $\sqrt s=1.96~\TeV$}

\affiliation{LAFEX, Centro Brasileiro de Pesquisas F\'{i}sicas, Rio de Janeiro, Brazil}
\affiliation{Universidade do Estado do Rio de Janeiro, Rio de Janeiro, Brazil}
\affiliation{Universidade Federal do ABC, Santo Andr\'e, Brazil}
\affiliation{University of Science and Technology of China, Hefei, People's Republic of China}
\affiliation{Universidad de los Andes, Bogot\'a, Colombia}
\affiliation{Charles University, Faculty of Mathematics and Physics, Center for Particle Physics, Prague, Czech Republic}
\affiliation{Czech Technical University in Prague, Prague, Czech Republic}
\affiliation{Institute of Physics, Academy of Sciences of the Czech Republic, Prague, Czech Republic}
\affiliation{Universidad San Francisco de Quito, Quito, Ecuador}
\affiliation{LPC, Universit\'e Blaise Pascal, CNRS/IN2P3, Clermont, France}
\affiliation{LPSC, Universit\'e Joseph Fourier Grenoble 1, CNRS/IN2P3, Institut National Polytechnique de Grenoble, Grenoble, France}
\affiliation{CPPM, Aix-Marseille Universit\'e, CNRS/IN2P3, Marseille, France}
\affiliation{LAL, Universit\'e Paris-Sud, CNRS/IN2P3, Orsay, France}
\affiliation{LPNHE, Universit\'es Paris VI and VII, CNRS/IN2P3, Paris, France}
\affiliation{CEA, Irfu, SPP, Saclay, France}
\affiliation{IPHC, Universit\'e de Strasbourg, CNRS/IN2P3, Strasbourg, France}
\affiliation{IPNL, Universit\'e Lyon 1, CNRS/IN2P3, Villeurbanne, France and Universit\'e de Lyon, Lyon, France}
\affiliation{III. Physikalisches Institut A, RWTH Aachen University, Aachen, Germany}
\affiliation{Physikalisches Institut, Universit\"at Freiburg, Freiburg, Germany}
\affiliation{II. Physikalisches Institut, Georg-August-Universit\"at G\"ottingen, G\"ottingen, Germany}
\affiliation{Institut f\"ur Physik, Universit\"at Mainz, Mainz, Germany}
\affiliation{Ludwig-Maximilians-Universit\"at M\"unchen, M\"unchen, Germany}
\affiliation{Panjab University, Chandigarh, India}
\affiliation{Delhi University, Delhi, India}
\affiliation{Tata Institute of Fundamental Research, Mumbai, India}
\affiliation{University College Dublin, Dublin, Ireland}
\affiliation{Korea Detector Laboratory, Korea University, Seoul, Korea}
\affiliation{CINVESTAV, Mexico City, Mexico}
\affiliation{Nikhef, Science Park, Amsterdam, the Netherlands}
\affiliation{Radboud University Nijmegen, Nijmegen, the Netherlands}
\affiliation{Joint Institute for Nuclear Research, Dubna, Russia}
\affiliation{Institute for Theoretical and Experimental Physics, Moscow, Russia}
\affiliation{Moscow State University, Moscow, Russia}
\affiliation{Institute for High Energy Physics, Protvino, Russia}
\affiliation{Petersburg Nuclear Physics Institute, St. Petersburg, Russia}
\affiliation{Instituci\'{o} Catalana de Recerca i Estudis Avan\c{c}ats (ICREA) and Institut de F\'{i}sica d'Altes Energies (IFAE), Barcelona, Spain}
\affiliation{Uppsala University, Uppsala, Sweden}
\affiliation{Taras Shevchenko National University of Kyiv, Kiev, Ukraine}
\affiliation{Lancaster University, Lancaster LA1 4YB, United Kingdom}
\affiliation{Imperial College London, London SW7 2AZ, United Kingdom}
\affiliation{The University of Manchester, Manchester M13 9PL, United Kingdom}
\affiliation{University of Arizona, Tucson, Arizona 85721, USA}
\affiliation{University of California Riverside, Riverside, California 92521, USA}
\affiliation{Florida State University, Tallahassee, Florida 32306, USA}
\affiliation{Fermi National Accelerator Laboratory, Batavia, Illinois 60510, USA}
\affiliation{University of Illinois at Chicago, Chicago, Illinois 60607, USA}
\affiliation{Northern Illinois University, DeKalb, Illinois 60115, USA}
\affiliation{Northwestern University, Evanston, Illinois 60208, USA}
\affiliation{Indiana University, Bloomington, Indiana 47405, USA}
\affiliation{Purdue University Calumet, Hammond, Indiana 46323, USA}
\affiliation{University of Notre Dame, Notre Dame, Indiana 46556, USA}
\affiliation{Iowa State University, Ames, Iowa 50011, USA}
\affiliation{University of Kansas, Lawrence, Kansas 66045, USA}
\affiliation{Louisiana Tech University, Ruston, Louisiana 71272, USA}
\affiliation{Northeastern University, Boston, Massachusetts 02115, USA}
\affiliation{University of Michigan, Ann Arbor, Michigan 48109, USA}
\affiliation{Michigan State University, East Lansing, Michigan 48824, USA}
\affiliation{University of Mississippi, University, Mississippi 38677, USA}
\affiliation{University of Nebraska, Lincoln, Nebraska 68588, USA}
\affiliation{Rutgers University, Piscataway, New Jersey 08855, USA}
\affiliation{Princeton University, Princeton, New Jersey 08544, USA}
\affiliation{State University of New York, Buffalo, New York 14260, USA}
\affiliation{University of Rochester, Rochester, New York 14627, USA}
\affiliation{State University of New York, Stony Brook, New York 11794, USA}
\affiliation{Brookhaven National Laboratory, Upton, New York 11973, USA}
\affiliation{Langston University, Langston, Oklahoma 73050, USA}
\affiliation{University of Oklahoma, Norman, Oklahoma 73019, USA}
\affiliation{Oklahoma State University, Stillwater, Oklahoma 74078, USA}
\affiliation{Brown University, Providence, Rhode Island 02912, USA}
\affiliation{University of Texas, Arlington, Texas 76019, USA}
\affiliation{Southern Methodist University, Dallas, Texas 75275, USA}
\affiliation{Rice University, Houston, Texas 77005, USA}
\affiliation{University of Virginia, Charlottesville, Virginia 22904, USA}
\affiliation{University of Washington, Seattle, Washington 98195, USA}
\author{V.M.~Abazov} \affiliation{Joint Institute for Nuclear Research, Dubna, Russia}
\author{B.~Abbott} \affiliation{University of Oklahoma, Norman, Oklahoma 73019, USA}
\author{B.S.~Acharya} \affiliation{Tata Institute of Fundamental Research, Mumbai, India}
\author{M.~Adams} \affiliation{University of Illinois at Chicago, Chicago, Illinois 60607, USA}
\author{T.~Adams} \affiliation{Florida State University, Tallahassee, Florida 32306, USA}
\author{J.P.~Agnew} \affiliation{The University of Manchester, Manchester M13 9PL, United Kingdom}
\author{G.D.~Alexeev} \affiliation{Joint Institute for Nuclear Research, Dubna, Russia}
\author{G.~Alkhazov} \affiliation{Petersburg Nuclear Physics Institute, St. Petersburg, Russia}
\author{A.~Alton$^{a}$} \affiliation{University of Michigan, Ann Arbor, Michigan 48109, USA}
\author{A.~Askew} \affiliation{Florida State University, Tallahassee, Florida 32306, USA}
% ...
%
%
\collaboration{The D0 Collaboration\footnote{with visitors from
$^{a}$Augustana College, Sioux Falls, SD, USA,
$^{b}$The University of Liverpool, Liverpool, UK,
$^{c}$DESY, Hamburg, Germany,
$^{d}$Universidad Michoacana de San Nicolas de Hidalgo, Morelia, Mexico
$^{e}$SLAC, Menlo Park, CA, USA,
$^{f}$University College London, London, UK,
$^{g}$Centro de Investigacion en Computacion - IPN, Mexico City, Mexico,
$^{h}$Universidade Estadual Paulista, S\~ao Paulo, Brazil,
$^{i}$Karlsruher Institut f\"ur Technologie (KIT) - Steinbuch Centre for Computing (SCC),
D-76128 Karlsruhe, Germany,
$^{j}$Office of Science, U.S. Department of Energy, Washington, D.C. 20585, USA,
$^{k}$American Association for the Advancement of Science, Washington, D.C. 20005, USA,
$^{l}$Kiev Institute for Nuclear Research, Kiev, Ukraine,
$^{m}$University of Maryland, College Park, Maryland 20742, USA
and
$^{n}$European Orgnaization for Nuclear Research (CERN), Geneva, Switzerland
}} \noaffiliation
\vskip 0.25cm

\date{January 15, 2015}
\begin{abstract}
We measure the ratio of cross sections,
$\sigma(p\bar{p}\rightarrow Z+2~b~\text{jets})$/$\sigma(p\bar{p}\rightarrow Z+\text{2~jets})$,
for associated production of a $Z$ boson with at least two jets with transverse momentum $p_T^{\rm jet} > 20$~GeV and pseudorapidity $|\eta^{\rm jet}| < 2.5$. 
This measurement uses data corresponding to an integrated luminosity of 9.7 fb$^{-1}$ collected by the \DO experiment in 
Run~II of Fermilab's Tevatron  \ppbar Collider at a center-of-mass energy 
of 1.96 TeV. The measured integrated
ratio of $0.0236\pm0.0032\left(\mbox{stat}\right)\pm0.0035\left(\mbox{syst}\right)$
is in agreement with predictions from next-to-leading-order perturbative QCD and the Monte Carlo event generators {\sc pythia} and {\sc alpgen}. 
\end{abstract}


\maketitle

Studies of $Z$ boson production in association 
with a bottom and an anti-bottom quark provide important tests of the predictions of
perturbative quantum chromodynamics (pQCD)~\cite{Campbell,Doreen}. 
A good theoretical description of this process is essential since it 
forms a major background for a variety of physics processes, including
standard model (SM) Higgs boson production in association with
a $Z$ boson, $ZH(H\rightarrow b\bar{b})$~\cite{zhllbb}, and
searches for supersymmetric partners of the $b$ quark~\cite{sbottom}.

The ratio of $Z+b~\text{jet}$ to $Z+\text{jet}$ production cross sections, 
for events with at least one jet, has been
previously measured by the CDF~\cite{CDFPaper,CDFPaperII} and \dzero\cite{Zb_PRL, Zb_PRD, Zb_PRD_2} collaborations using Run~II data.
The ATLAS~\cite{atlas} and CMS~\cite{cms} collaborations have also 
studied $Z+b~\text{jet}$ production at $\sqrt{s} = 7$~TeV.

This article presents the ratio of $Z+2~b~\text{jets}$ to $Z+\text{2~jets}$ inclusive production cross sections and is an extension of the previous \dzero measurements utilizing similar event selections. 
The measurement of the ratio benefits from the cancellation of many systematic uncertainties, such as
the uncertainty in luminosity and those related to lepton and jet identification, allowing a more precise
comparison with theory. The remaining systematic uncertainties arise from the differences
between $b$ jets and light jets. In the following, light-quark flavor ($u$, $d$, $s$) and gluon jets are referred to as ``light jets''.
 The $Z+2~b~\text{jet}$ production cross sections have been 
measured at CMS ~\cite{CMS_2b} and ATLAS~\cite{Atlas_2b} at $\sqrt{s} = 7$~TeV.
The current measurement is based on the complete Run~II
data sample collected by the D0 experiment  at the
Fermilab Tevatron \ppbar collider at a center-of-mass
energy of $\sqrt{s}=1.96$~TeV, and corresponds to an integrated luminosity of 9.7 fb$^{-1}$.

We first briefly describe the main components of the D0 Run II detector
\cite{d0det,d0det2} relevant to this analysis. The D0 detector has a central tracking system consisting of a
silicon microstrip tracker (SMT) \cite{layer0} and a central fiber tracker (CFT),
both located within a 1.9~T superconducting solenoidal
magnet, with designs optimized for tracking and
vertexing at pseudorapidities $|\eta_{\text {det}}|<3$ and $|\eta_{\text {det}}|<2.5$, respectively~\cite{coord}.
A liquid argon and uranium calorimeter has a
central section (CC) covering pseudorapidities $|\eta_{\text {det}}| \lesssim 1.1$, and two end calorimeters (EC) that extend coverage
to $|\eta_{\text {det}}|\approx 4.2$, with all three housed in separate
cryostats~\cite{calopaper}. An outer muon system, at $|\eta_{\text {det}}|<2$,
consists of a layer of tracking detectors and scintillation 
counters in front of 1.8~T toroids, followed by two similar layers
after the toroids.
Luminosity is measured using plastic scintillator
arrays located in front of the EC cryostats. The trigger and data
acquisition systems are designed to accommodate the high instantaneous luminosities
of Run~II.

This analysis relies on all components of the \dzero detector: tracking systems, the liquid-argon sampling calorimeter, muon system, 
and the ability to identify secondary vertices~\cite{d0det}. 
The SMT allows for precise reconstruction 
of the primary \ppbar interaction vertex 
and secondary vertices ~\cite{coord,PV}. It also enables  
an accurate determination of the impact parameter, defined as
the distance of closest approach of a track to the primary interaction vertex in the $x$-$y$ plane. The impact
parameter measurements of tracks, along with reconstructed secondary vertices, 
are important inputs to the $b$-jet tagging algorithm.

Events containing $Z$ bosons decaying to $\mu\mu$ or $ee$ are collected using triggers based on single electrons or muons.
For the off-line selection requirements discussed below, the triggers have an
efficiency of approximately 100\% for $Z\rightarrow ee$ and more
than 78\% for $Z\rightarrow\mu\mu$ decays depending on the transverse momentum of the muon.
The $Z+\text{2~jet}$ sample requires the presence of
at least two jets in the event, while the $Z+2~b~\text{jet}$ sample
requires at least two $b$-jet candidates, selected using a
$b$-tagging algorithm~\cite{bid}.

An event is selected if it contains a \ppbar interaction vertex, reconstructed from 
at least three tracks, located within 60 cm of the center of 
the \DO detector along the beam axis.
The selected events must also contain a $Z$ boson candidate with a
dilepton invariant mass $70~<M_{\ell\ell}<110~\GeVe$.

Dielectron ($ee$) events are required to have two electrons of transverse 
momentum ($p_{T}$) greater than $15~\GeVe$ identified through electromagnetic (EM) showers in the calorimeter. 
The showers must have more than 97\% of their energy deposited in the EM
calorimeter, be isolated from other energy depositions, and have transverse 
and longitudinal energy profiles consistent with that expected for  electrons. 
At least one electron must be identified in the CC, with $|\eta_{\text {det}}|<1.1$,
and a second electron either in the CC or the EC, $1.5<|\eta_{\text {det}}|<2.5$.
Electron candidates in the CC are required to match central tracks 
or have a pattern of hits consistent with the passage of an electron through the 
central tracker. Electrons in the ECs are not required to have a track matched to them due to deteriorating tracking coverage for $|\eta_{\text {det}}|>2$.
Due to the lack of track requirement in EC regions we do not apply any opposite sign requirement for the dielectron events.

The dimuon ($\mu\mu$) event selection requires two oppositely charged 
muons detected in the muon system that are matched to reconstructed tracks in the central tracker with \pt$>15~\GeVe$ and $|\eta_{\text {det}}|<2$. These muons must pass a combined 
tracking and calorimeter isolation requirement discussed in detail in Ref.~\cite{zhllbb}.
Muons originating from cosmic rays are rejected by 
applying timing criteria using the hits in the scintillation counters and by 
limiting the measured displacement of the muon track with respect to the \ppbar interaction vertex~\cite{muonid}.

A total of about 1.2 million $Z$ boson candidate events 
are retained in the combined $ee$ and $\mu\mu$ channels with the above
lepton selection criteria. The $Z+\text{2~jet}$ sample is then selected by requiring at 
least two jets in the event with \ptj~$>20~\GeVe$ and $|\etaj|<2.5$.
Jets are reconstructed from energy 
deposits in the calorimeter using an iterative midpoint cone algorithm~\cite{RunIIcone} 
with a cone of radius
$\Delta R =  \sqrt{(\Delta\varphi)^{2}+(\Delta y)^{2}}= 0.5$ where
$\varphi$ is the azimuthal angle and $y$ is the rapidity.
Jet energy is corrected for detector response, the presence of noise
and multiple $p\bar{p}$ interactions. We also correct the jet energy for the energy of those particles within the reconstruction cone that is deposited in the calorimeter outside the cone (and vice versa)~\cite{JES}. 

To suppress background from top-antitop quark ($t\bar{t}$) production, events are rejected if
the missing transverse energy is larger than $60~\GeVe$, reducing 
the $t\bar{t}$ contamination by a factor of two.
These selection criteria 
retain an inclusive sample of 20,950 $Z+\text{2~jet}$ event candidates in 
the combined $ee$ and $\mu\mu$ channels.

Processes such as diboson ($WW$, $WZ$, $ZZ$) production can
contribute to the background when 
two leptons are reconstructed in the final state.
Inclusive diboson production is simulated with the
{\sc pythia}~\cite{pythia} Monte Carlo (MC) event generator. The $Z+\text{jet}$, including heavy flavor jets, and $t\bar{t}$ events are modeled by
{\sc alpgen}~\cite{alpgen}, which generates hard sub-processes including higher 
order QCD tree level matrix elements, interfaced with
{\sc pythia} for parton showering and hadronization.
The {\sc CTEQ6L1}~\cite{cteq6} 
parton distribution functions (PDFs) are used in all simulations.
The cross sections of the simulated samples are then scaled to the corresponding 
higher-order theoretical calculations. 
For the diboson and $Z+\text{2~jet}$ processes, 
including the $Z + b\bar{b}$ signal process and $Z + c\bar{c}$
production, next-to-leading order (NLO) cross section
predictions are taken from {\sc mcfm}~\cite{diboson}.
The $t\bar{t}$ cross section is determined from NLO+NNLL (next-to-next-leading log)
calculations~\cite{ttbar}. 
To improve the modeling of the \pt distribution of the $Z$ boson, simulated 
$Z+\text{2~jet}$ events are also reweighted to be consistent with the measured \pt 
spectrum of $Z$ bosons observed in data~\cite{zpt}. 

These generated samples are processed 
through a detailed detector simulation based on {\sc geant}~\cite{geant}. 
To model the effects of detector noise 
and pile-up events, collider data from random beam crossings with the same instantaneous luminosity distribution as for data are 
superimposed on simulated events.
These events are then reconstructed using the same algorithms as used for data.
Scale factors, determined from data using independent samples, are applied to 
account for differences in reconstruction efficiency between data and simulation. 
The energies of simulated jets are corrected, based on their flavor, to 
reproduce the resolution and energy scale observed in data~\cite{JES}.  

The background contribution from multijet events, 
in which jets are misidentified as leptons,
is evaluated from data. This is performed using a multijet-enriched sample of
events that pass all selection criteria except for some of the lepton quality
requirements. In the case of electrons, the multijet sample is obtained by
inverting the shower shape requirement and relaxing other electron
identification criteria, while for the muon channel, the multijet sample consists
of events with muon candidates that fail the isolation requirements. 
The normalization of the multijet background is
determined from a simultaneous fit to the dilepton invariant
mass distributions in different jet multiplicity bins.

\begin{figure}
\includegraphics[width=0.46\textwidth]{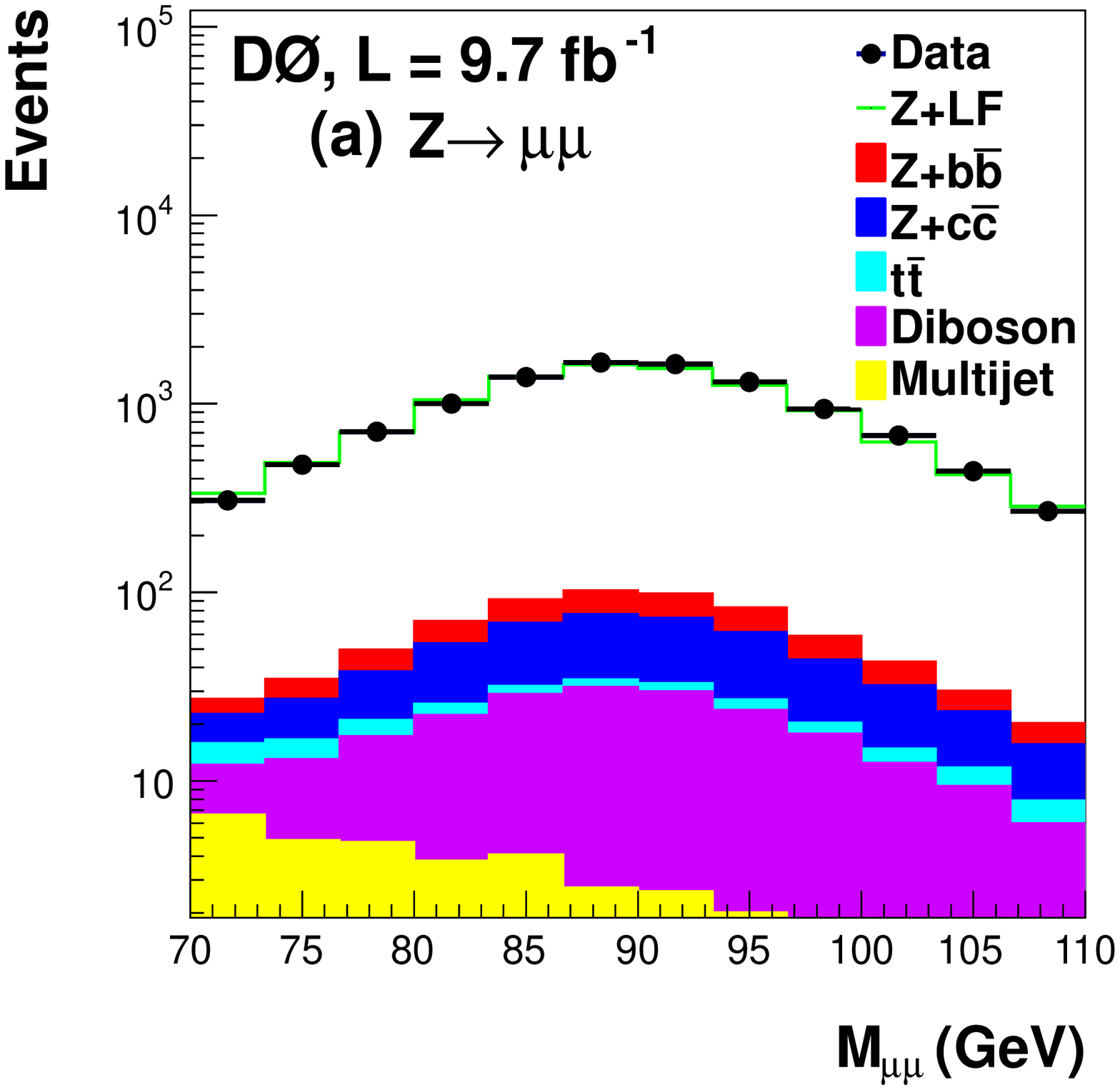}
\includegraphics[width=0.46\textwidth]{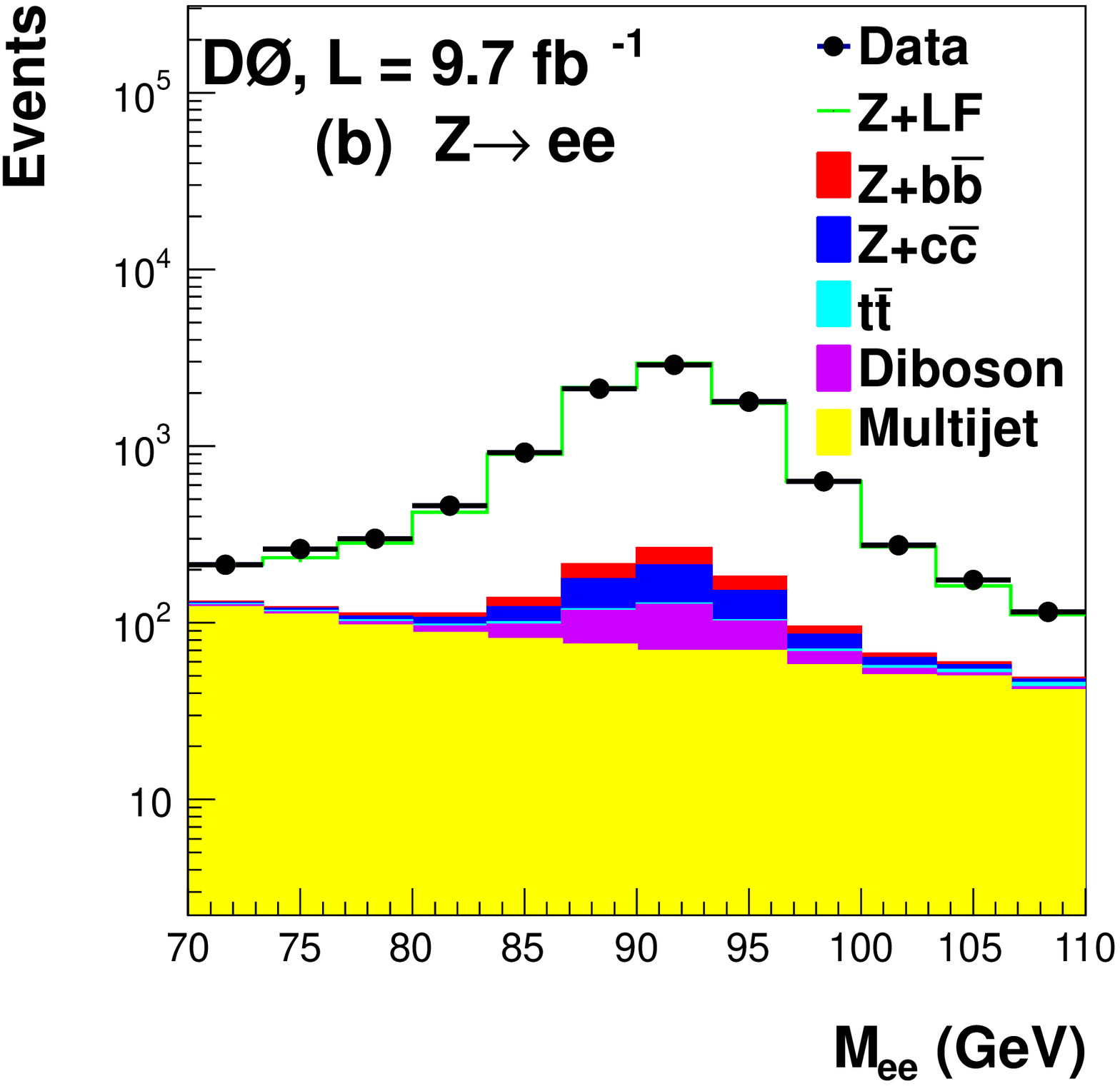}
\caption{\label{fig:Zm_presel}(color online) The invariant mass in (a) $Z\rightarrow \mu\mu$ and (b) $Z\rightarrow ee$ channels for data and background in events with a $Z$ boson
candidate and at least two jets before $b$ tagging is applied.}
\end{figure}

\begin{figure}
\includegraphics[width=0.44\textwidth]{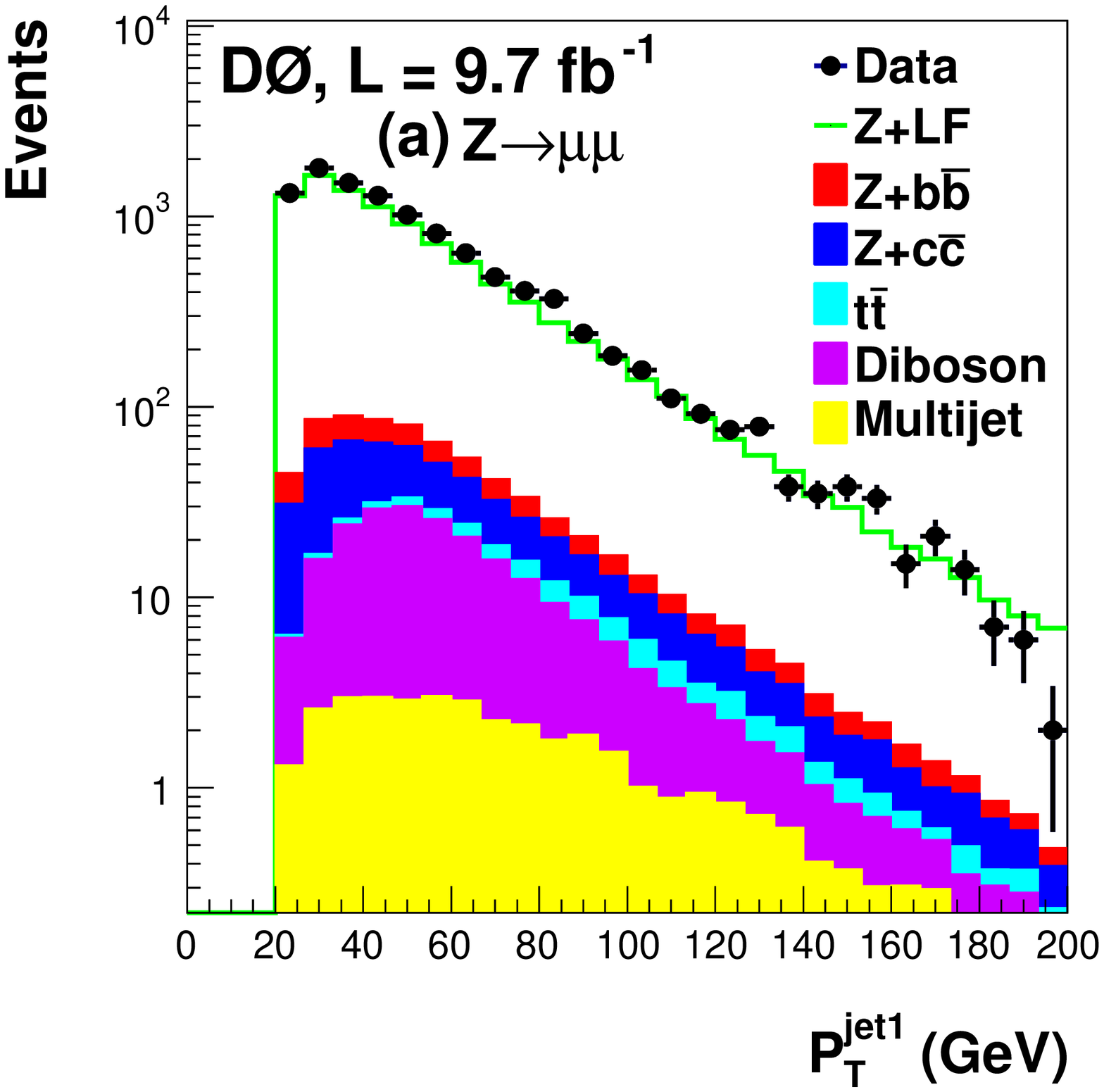}
\includegraphics[width=0.46\textwidth]{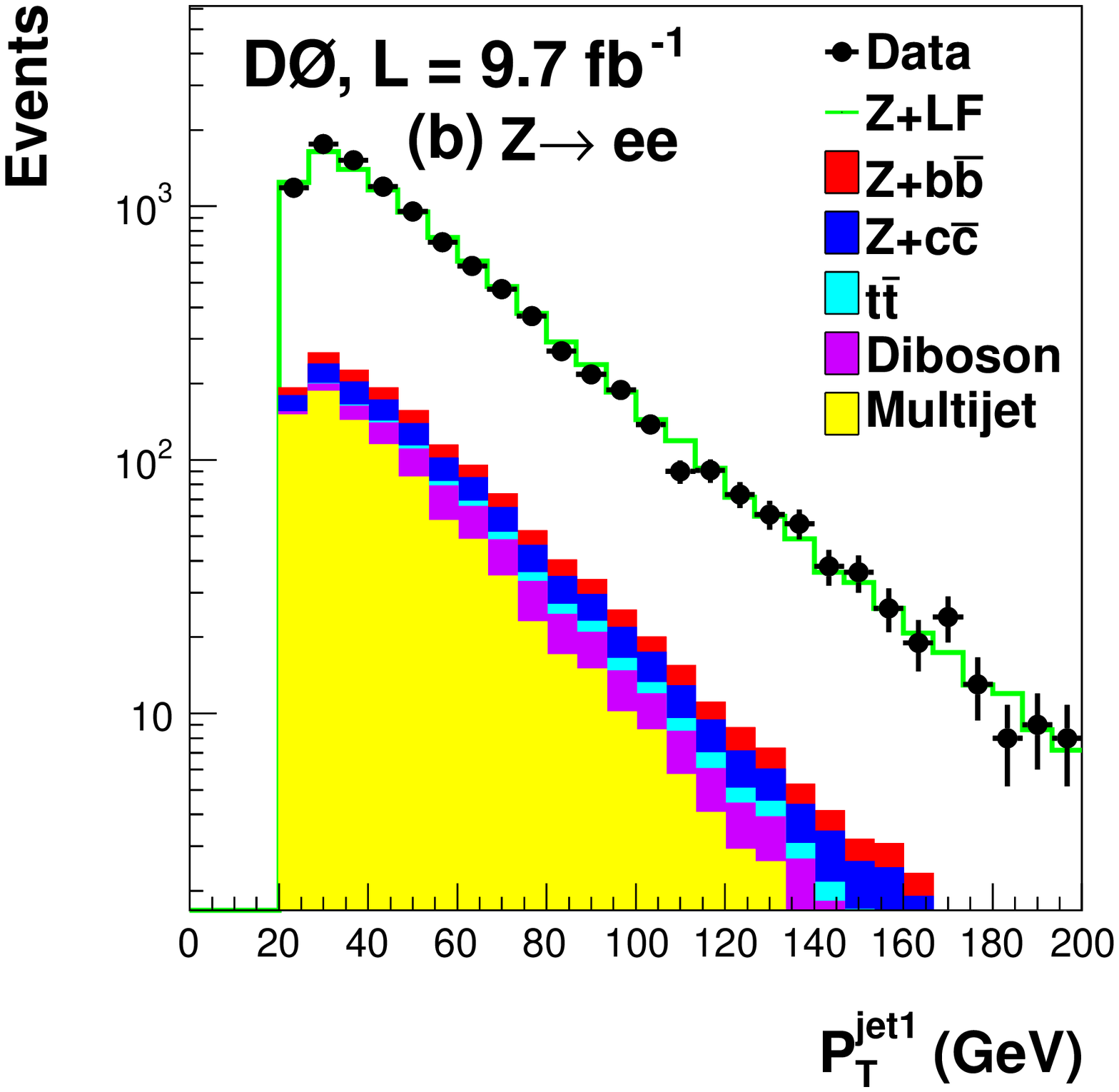}
\caption{\label{fig:jetpT_presel}(color online) The leading jet $p_{T}$ in the (a) $Z\rightarrow \mu\mu$ and (b) $Z\rightarrow ee$ channels for data and background in events with a $Z$ boson
candidate and at least two jets before $b$ tagging is applied.}
\end{figure}

Figures~\ref{fig:Zm_presel} and ~\ref{fig:jetpT_presel} show the dilepton invariant mass and leading jet $p_{T}$ distributions
in data compared to the expectations from various processes.
The dominant contribution comes from $Z+$light jet production.
The non-$Z+\text{jet}$ background fraction in the $ee$ channel is about 15\%,
and is dominated by multijet production. The muon channel 
has a higher purity with a background fraction of about 7\%.

This analysis employs a two-step procedure to determine the $b$-quark 
content of jets in the selected events. First, a $b$-tagging algorithm
is applied to jets to select a sample of $Z+\text{2~jet}$ events
that is enriched in heavy flavor jets. After $b$ tagging, the relative light, $c$, 
and $b$-quark content is extracted by fitting
templates built from a dedicated
discriminant that provides an optimized separation between the three components. 

Jets considered for $b$-jet tagging are subject to a preselection requirement,
called taggability, to decouple the intrinsic performance of the $b$-jet tagging
algorithm from effects related to the track reconstruction efficiency. 
For this purpose, the jet is required to have at least two associated
tracks with \pt$>0.5~\GeVe$, the leading track must have
\pt$>1~\GeVe$, and each track must have at least one
SMT hit. This requirement has a typical efficiency of 90\% per jet.

The $b$-jet tagging algorithm is based on a 
multivariate analysis (MVA) technique~\cite{MVA}. This algorithm, MVA$_{bl}$, 
discriminates $b$ jets from light-flavor jets
utilizing the relatively long lifetime of the $b$ hadrons when compared 
to their lighter counterparts~\cite{bid}. Events with at least two jets tagged 
by this algorithm are considered.

The MVA$_{bl}$ discriminant combines various properties of the jet and associated tracks
to create a continuous output that tends towards unity for $b$ jets and zero for
light jets. 
Inputs include the number of secondary vertices and the 
charge track multiplicity, invariant mass of the secondary vertex ($M_{\text{SV}}$), decay length 
and impact parameter of secondary vertices, the multiplicity of charged 
tracks associated with them, and the Jet Lifetime Probability (JLIP), which
is the probability that tracks associated with the jet originate from the \ppbar interaction vertex~\cite{bid}.
Events are retained for further analysis if they contain at least two jets with an MVA$_{bl}$
output greater than 0.15.
After these requirements, 241 $Z+\text{2~jet}$ events are
selected with at least two $b$-tagged jets, where only the 
two highest \pt tagged jets
are examined in the analysis and the electron and muon channels are combined. The efficiency for tagging
two $b$ jets in data is 33\%. In the MC correction factors are applied to account for differences with data~\cite{bid}. The background 
contamination from diboson, multijet, and top production after $b$-tagging, 
for the electron and muon channels combined are 8\%, 2\% and 15\%  respectively. 

To determine the fraction of events with $2~b$ jets, a dedicated discriminant, \mjle, is employed~\cite{Zb_PRD,Wb}. 
It is a combination of the two most discriminating MVA$_{bl}$ inputs, 
$M_{\text{SV}}$ and JLIP: $\mjle = 0.5\times(M_{\rm SV}/(5~{\rm GeV})-\ln({\rm JLIP})/20)$.

\begin{figure}
 \includegraphics[width=0.45\textwidth]{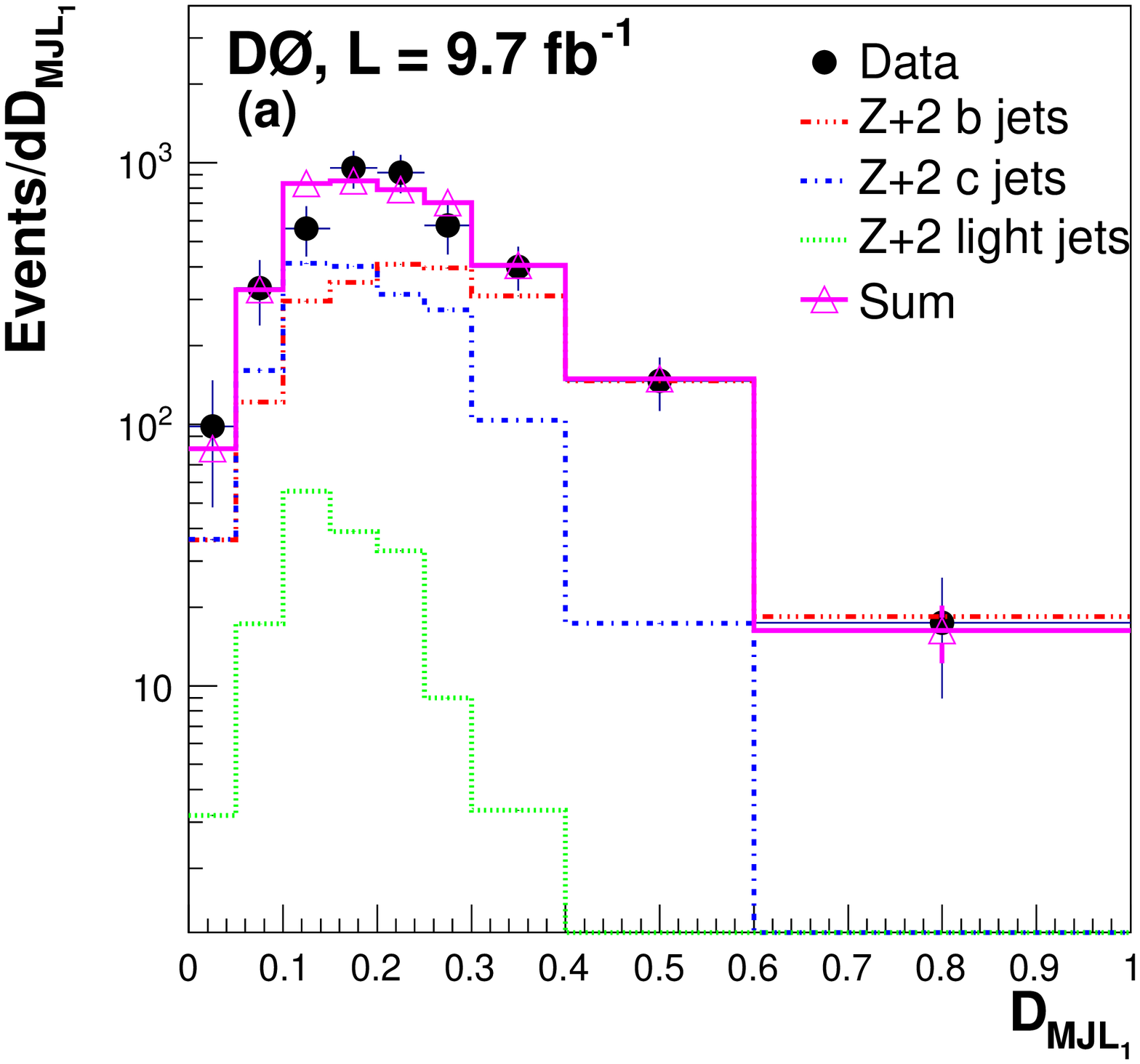}
 \includegraphics[width=0.45\textwidth]{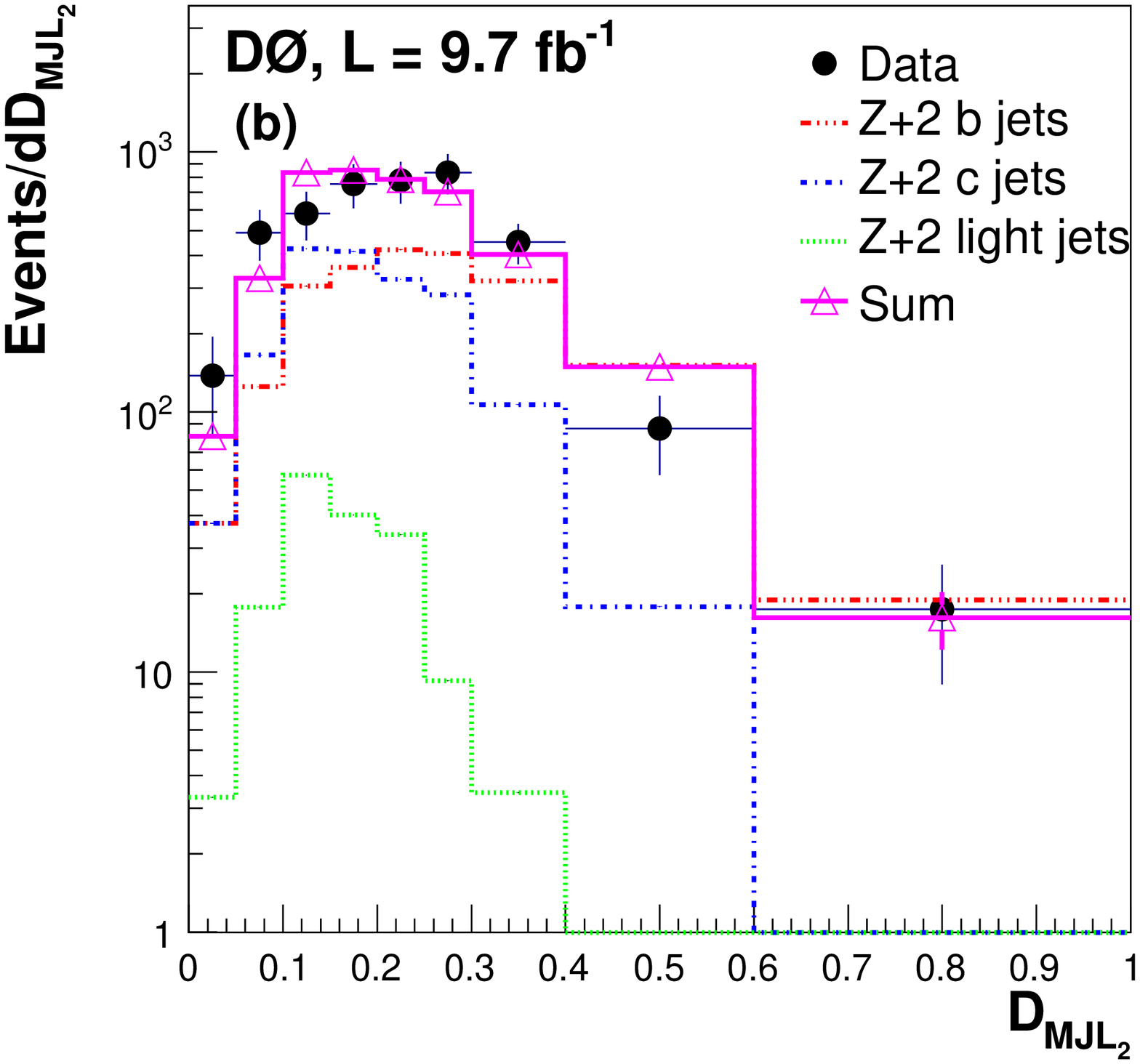}
\caption{\label{fig:channel_fit} (color online) 
The one dimensional projection onto (a) the highest-$p_T$ jet and (b) the second highest-$p_T$ jet $D_{\rm MJL}$ axis of the 2D fit. The distributions of the $b$, $c$,
and light jets are normalized by the fractions found from the fit.}
\end{figure}
To measure the fraction of events with different jet flavors
in the selected sample, we count the number of events as a function
of the $D_{\rm MJL}$ of the two leading jets $N(D_{\rm MJL_1},D_{\rm MJL_2})$ and then perform a two dimensional
binned maximum likelihood fit to that distribution. The data sample with two heavy-flavor-tagged jets is fitted to templates
 consisting mainly of $2~b$-jet, $2~c$-jet, and light
flavor jet events, as obtained from {\sc alpgen+pythia} simulated samples. We also compared the shapes of the templates from {\sc SHERPA} simulated samples 
and found the templates to be consistent for the two models. 
Before the fit, all non-$Z+jet$ background 
contributions, estimated from simulated samples after the MVA$_{bl}$ requirement, 
are subtracted from the data leaving 180 $Z+\text{2~jet}$
events in the combined $ee$ and $\mu\mu$ channel.
Next, we measure the jet-flavor fractions in the dielectron and dimuon
samples combined, yielding the $2~b$ jet flavor fraction ($f_{bb}$) of 
$0.64\pm0.08~(\mbox{stat.})$ and the $2~c$ jet flavor fraction of $0.32\pm0.08~(\mbox{stat.})$. 
 Figure~\ref{fig:channel_fit} shows the one dimensional projection onto the highest-$p_T$ jet and the second highest-$p_T$ jet $D_{\rm MJL}$ axis of the 2D fit.

The fraction of $2~b$ jets measured in the heavy flavor enriched sample 
is combined with the corresponding event acceptances 
to determine the ratio, $R$, of the cross sections.

\begin{eqnarray}
R=\frac{\sigma(Z+2~b~\text{jets})}{\sigma(Z+\text{2~jets})}=\frac{N_{bb}\, f_{bb}}{N_{\text{incl}}\,\epsilon_{tag}^{bb}}\times\frac{\mathcal{A}_{\text{incl}}}{\mathcal{A}_{bb}}
\label{eq:master}
\end{eqnarray}

\noindent where  $N_{\text{incl}}$ is the total number of $Z+\text{2~jet}$ events
before the tagging requirements, $N_{bb}$ is the number of $Z+\text{2~jet}$
events used in the \mjl fit, $f_{bb}$ is the extracted $2~b$ jet
fraction, and $\epsilon_{tag}^{bb}$ is the overall selection efficiency of \mjl 
for $2~b$ jets that combines the efficiencies for taggability and MVA$_{bl}$
discriminant. Both $N_{\text{incl}}$ and $N_{bb}$
correspond to the number of events that remain after the contributions 
from non-$Z+\text{jets}$ processes have been subtracted from the data.
	
The pseudorapidity acceptance for electrons and muons is different.  
In order to quote a combined ratio for the two channels, we correct to a common lepton acceptance as follows.
The detector acceptances for the inclusive jet sample
and $2~b$ jets are determined from MC simulations in the kinematic region that
satisfies the $p_T$ and $\eta$ requirements for leptons and jets. For the $\mathcal{A}_{bb}$ and $\mathcal{A}_{\text{incl}}$ calculations, 
we apply selections for both the electron and muon channels for the fiducial region for the events with two jets and two leptons defined as:
\begin{eqnarray}
p_T^{\rm jet} > 20~{\rm GeV}~{\rm and}~|\eta^{\rm jet}| < 2.5, \nonumber \\
p_T^\ell > 15~{\rm GeV}~{\rm and}~|\eta^\ell| < 2.
\label{eq:fiducial_cuts}
\end{eqnarray}

The resulting ratio of the two acceptances is measured to be
$\mathcal{A}_{\text{incl}}/\mathcal{A}_{bb}=1.09\pm0.02~(\mbox{stat})$.

Using Eq.~(\ref{eq:master}), we obtain the ratio of the $Z+2~b~\text{jet}$ cross section to the
inclusive $Z+\text{2~jet}$ cross section in the combined $\mu\mu$ and $ee$ channel
to be $0.0236\pm0.0032~(\text{stat})$.

Several systematic uncertainties cancel when the ratio 
\ratio~is measured. 
These include uncertainties
on the luminosity measurement, lepton trigger efficiency, and lepton and 
jet reconstruction efficiencies. The remaining uncertainties 
are estimated separately for the integrated result. The largest systematic uncertainty of 13.7\% comes from the uncertainty on the 
shape of the \mjl templates used 
in the fit including that due to MC statistics of the samples used to construct the templates. The shape of the templates may be affected by the choice 
of the $b$ quark fragmentation function~\cite{topmass},
the background estimation, the difference in 
the shape of the light jet MC template and a template derived from a light jet enriched dijet data sample, 
and the composition of the charm states used to determine the charm template shape~\cite{Zb_PRD}. It also includes uncertainties on production rates of different hadrons and uncertainties on branching fractions. These effects are evaluated by varying
the central values by the corresponding uncertainties, one at a time. 
The entire analysis chain is checked for possible biases
using a MC closure test and no significant deviations are observed.
The next largest systematic uncertainty of 5.5\% is due to the $b$-jet identification efficiency.
The uncertainty on $b$-jet energy calibration is 2.6\%; it comprises the uncertainties on the jet energy
resolution and the jet energy scale.
For the integrated ratio measurement, these uncertainties, 
when summed in quadrature, result in a
total systematic uncertainty of 14.9\%. 
For the integrated ratio we obtain
\begin{eqnarray}
R = 0.0236\pm0.0032\left(\text{stat}\right)\pm0.0035\left(\mbox{syst}\right).
\end{eqnarray}
To check the stability of the result, the ratio is remeasured using a looser(tighter) MVA$_{bl}$ selection with the lower limit on the
MVA$_{bl}$ output of $>0.10$($>0.225$). The looser selection provides increased 
data statistics and the tighter selection yields a $2~b$ enriched sample. The new and default ratios are found to be in agreement within uncertainties of about 4\%.

To validate the $t{\bar t}$ background estimation, we reduce the contribution of $t{\bar t}$ events by rejecting 
events where the scalar sum of all jet $p_T$ values is more than 130 GeV.
This selection reduces the $t{\bar t}$ fraction by an additional factor of $2$ with a signal efficiency of $80\%$.
The new and default ratios are found to be in agreement within systematic uncertainties.

In Table \ref{tab:res_b}, we present the ratio of integrated cross sections,
$\sigma(p\bar{p}\rightarrow Z+2~b~\text{jet})$/$\sigma(p\bar{p}\rightarrow Z+\text{2~jet})$, in the fiducial region defined in Eq.~(\ref{eq:fiducial_cuts}).
The ratio is compared to predictions from NLO QCD calculations and two MC generators, {\sc pythia} and {\sc alpgen}.
The NLO predictions use the MSTW2008 PDF set~\cite{mstw} 
 using \textsc{mcfm} with central values of renormalization and fragmentation scales $\mu_{r}=\mu_{f}=M_Z$.
Uncertainties are estimated by varying $\mu_{r}$ and $\mu_{f}$ together by a factor of two, and  are about 15\%.
{\sc alpgen} generates multi-parton final states using tree-level
matrix elements. When interfaced with {\sc pythia}, it
employs an MLM scheme~\cite{MLM} to match matrix
element partons with those after showering in {\sc pythia}, 
resulting in an improvement over leading-logarithmic
accuracy. The measured ratio is in reasonable agreement with MCFM NLO calculations considering the uncertainties on the data and theory.

\begin{table*}
\begin{center}
\caption{The ratio of integrated cross sections,
$\sigma(p\bar{p}\rightarrow Z+2~b~\text{jet})$/$\sigma(p\bar{p}\rightarrow Z+\text{2~jet})$ together with statistical uncertainties ($\delta_{\text{stat}}$) and  total systematic uncertainties ($\delta_{\text{syst}}$).
The column $\delta_{\rm tot}$ shows the total experimental uncertainty obtained by adding $\delta_{\text{stat}}$
and $\delta_{\text{syst}}$ in quadrature.
The last three columns show theoretical predictions obtained using NLO QCD with scale uncertainties and two MC event generators,
{\sc pythia} and {\sc alpgen}.
}
\label{tab:res_b}
\renewcommand{\arraystretch}{2}
\begin{tabular}{ccccc} \hline

 \multicolumn{3}{c}{~~~~~~~~~~~~~~~~~~~~~~~~~~~~$\sigma(p\bar{p}\rightarrow Z+2~b~\text{jet})$/$\sigma(p\bar{p}\rightarrow Z+\text{2~jet})$ } \\\cline{1-5}
 Data $\pm \delta_{\rm stat}\pm \delta_{\rm syst}$ & $\delta_{\rm tot}$ & ~~~{\sc nlo qcd(mstw)}~~~ & ~~~{\sc pythia}~~~ & ~~~{\sc alpgen}~~~
\\\hline

$(2.36  \pm 0.32 \pm 0.35)\times 10^{-2}$ & 0.47$\times 10^{-2}$ & $(1.76\pm 0.26)\times 10^{-2}$  & 2.42$\times 10^{-2}$ &2.21$\times 10^{-2}$           
\\\hline
\end{tabular}
\end{center}
\end{table*}

In summary, we report the measurement at the Tevatron of the ratio of integrated cross sections,
$\sigma(p\bar{p}\rightarrow Z+2~b~\text{jet})$/$\sigma(p\bar{p}\rightarrow Z+\text{2~jet})$,
for events with $Z\rightarrow\ell\ell$ in a restricted phase space of leptons with $p_T^{\ell} > 15$~GeV,
$|\eta^{\ell}| < 2.0$ and with two jets limited to $p_T^{\rm jet} > 20$~GeV and $|\eta^{\rm jet}| < 2.5$. Measurements are based on
the full data sample collected by the \dzero experiment in Run~II of the Tevatron,
corresponding to an integrated luminosity of 9.7 fb$^{-1}$ at
a center-of-mass energy of 1.96 TeV. The measured integrated
ratio of $0.0236\pm0.0032\left(\mbox{stat}\right)\pm0.0035\left(\mbox{syst}\right)$
is in agreement with the theoretical predictions within uncertainties. 

%

We thank John Campbell and Doreen Wackeroth for
valuable discussions, and the staffs at Fermilab and collaborating institutions, and acknowledge support from the
Department of Energy and National Science Foundation (United States of America);
Alternative Energies and Atomic Energy Commission and
National Center for Scientific Research/National Institute of Nuclear and Particle Physics  (France);
Ministry of Education and Science of the Russian Federation,
National Research Center ``Kurchatov Institute" of the Russian Federation, and
Russian Foundation for Basic Research  (Russia);
National Council for the Development of Science and Technology and
Carlos Chagas Filho Foundation for the Support of Research in the State of Rio de Janeiro (Brazil);
Department of Atomic Energy and Department of Science and Technology (India);
Administrative Department of Science, Technology and Innovation (Colombia);
National Council of Science and Technology (Mexico);
National Research Foundation of Korea (Korea);
Foundation for Fundamental Research on Matter (The Netherlands);
Science and Technology Facilities Council and The Royal Society (United Kingdom);
Ministry of Education, Youth and Sports (Czech Republic);
Bundesministerium f\"{u}r Bildung und Forschung (Federal Ministry of Education and Research) and
Deutsche Forschungsgemeinschaft (German Research Foundation) (Germany);
Science Foundation Ireland (Ireland);
Swedish Research Council (Sweden);
China Academy of Sciences and National Natural Science Foundation of China (China);
and
Ministry of Education and Science of Ukraine (Ukraine).


\begin{thebibliography}{10}
\bibitem{coord}
We use a standard right-handed coordinate system. The nominal collision point is the center of the detector with coordinates $(0,0,0)$.
The direction of the proton beam is the positive $+z$ axis. The $+x$ axis is horizontal, pointing away from the center of the Tevatron ring.
The $+y$ axis points vertically upwards. The polar angle, $\theta$, is defined such that $\theta = 0 $ is the $+z$ direction.
The rapidity is defined as $y = -\ln \left[ (E+p_z)/(E-p_z)\right]$, where $E$ is the energy and $p_z$ is the momentum component along
the proton beam direction. Pseudorapidity is defined as $\eta = - \ln(\tan \frac{\theta}{2})$.
$\varphi$ is defined as the azimuthal angle in the plane transverse to the proton beam direction. Also, $\eta_{\text {det}}$ and $\phi_{\text {det}}$ are the pseudorapidity and the azimuthal angle
measured with respect to the center of the detector.
\bibitem{PV} The primary \ppbar interaction vertex is that found to be
the most likely collision point, among possibly several
collisions within a specific beam crossing, from which 
selected objects emanate. The algorithm for defining it
can be found in~\cite{bid}.
\end{thebibliography}
\end{document}